\documentclass[showpacs,preprintnumbers,amsmath,amssymb]{revtex4}
\usepackage{graphicx}
\usepackage{dcolumn}
\usepackage{bm}
\usepackage{amsmath}
\usepackage{multirow}
\usepackage{tikz}

\usepackage{./mw}

\begin{document}

\preprint{}

\title{Entropy and diffuse scattering: comparison of NbTiVZr and CrMoNbV}

\author{M. Widom}
\affiliation{
Department of Physics, Carnegie Mellon University,
Pittsburgh, PA  15213
}

\date{\today}
\begin{abstract}
The chemical disorder intrinsic to high entropy alloys inevitably creates diffuse scattering in their x-ray or neutron diffraction patterns.  Through first principles hybrid Monte Carlo/molecular dynamics simulations of two BCC high entropy alloy forming compounds, CrMoNbV and NbTiVZr, we identify the contributions of chemical disorder, atomic size and thermal fluctuations to the diffuse scattering. As a side benefit, we evaluate the reduction in entropy due to pair correlations within the framework of the cluster variation method. Finally, we note that the preference of Ti and Zr for HCP structures at low temperature leads to a mechanical instability reducing the local BCC character of NbTiVZr, while preserving global BCC symmetry.
\end{abstract}

\maketitle

\section{Introduction}
While the Bragg peaks of diffraction patterns reveal the global symmetries and average structures of crystals, diffuse scattering intensity located between the peaks contains information about the deviations from the average.  Many factors create diffuse scattering: chemical disorder creates smooth diffuse background due to contrasting scattering form factors; variation in atomic sizes and thermal motion (phonons) leads to diffuse wings that diverge in intensity near the Bragg peaks each with its own characteristic angular variation; a variety of extrinsic defects each with their own diffuse signature.

Because each type of disorder creates diffuse intensity of specific and quantifiable intensity patterns, a great deal of physical information can be extracted from diffuse scattering experiments.  Disorder in atomic form factors creates a uniform diffuse background, but short-range chemical order can cause the background to concentrate in specific regions between or away from Bragg peaks depending on whether the interatomic interactions favor local association of like or unlike chemical species.  Thermal diffuse scattering is dominated by long wavelength phonons and thus the bulk, shear and other elastic moduli can be extracted from its angular variation.  Atomic size scattering arises from long-range strain fields caused by mismatch of an atoms size to its environment. Its strength depends on the mismatch and on particular combinations of elastic constants.

We explore the diffuse scattering patterns of two high entropy alloys by means of computer simulation. The compounds NbTiVZr~\cite{Senkov13} and CrMoNbV~\cite{Widom15} are drawn from refractory metals of, respectively, the fourth and fifth, and the fifth and sixth columns of the periodic table. Thus the atomic form factors and the contrasts among the form factors are similar for the two cases.  Because both belong to squares of the periodic table~\cite{Widom13,Widom15}, we anticipate the greatest contrast in size to lie along the positive diagonals of the squares, namely Zr-V and Nb-Cr.  One interesting point about these alloy systems is that all constituent elements of CrMoNbV individually form body centered cubic (BCC) lattices, while two of the elements (Zr and Ti) of NbTiVZr are BCC only at high temperature but prefer hexagonal (HEX) lattices at low temperature. Indeed, the BCC structures of these elements are mechanically unstable at low temperature.

In order to efficiently explore the configurational ensemble in the solid state, we utilize a hybrid MC/MD method that alternates Monte Carlo swapping of atomic species with conventional molecular dynamics~\cite{Widom13}.  A side benefit of this simulation method is that we obtain partial pair correlation functions of the alloy systems.  From these correlation functions we can see the impact of atomic size differentials and preferences in short-range chemical order. Remarkably, we see a loss of short-range BCC geometry in the case of NbTiVZr.  This results in stronger diffuse scattering in NbTiVZr than in CrMoNbV.  The existence of short-range chemical order requires that the actual configurational entropy lie below its ideal value of $k_B\log{(4)}$.  We apply the formalism of the cluster variation method (CVM) to estimate the actual entropy (at T=1200K) of NbTiVZr as $k_B\log{(3.93)}$ and CrMoNbV as $k_B\log{(3.76)}$.

\section{Results}
\subsection{Swap rates and pair correlations}

Using hybrid MC/MD we pre-anneal structures of $N=128$ atoms (32 of each species) at T=1800K and then 1500K prior to lengthy data collection runs at 1200K.  Our data collection runs extend for 15ps of molecular dynamics and 1500 attempted swaps for NbTiVZr, and 28ps of MD and 2800 attempted swaps for CrMoNbV.  Acceptance rates for swaps indicate the similarity of chemical species in size and bonding preferences.  Characteristically, the swap rates are lowest across the positive diagonal of periodic table squares~\cite{Widom15} (i.e. Nb with Cr, and Zr with V) and highest along the negative diagonal (i.e. V with Mo, and ti with Nb).  The relatively large acceptance rates for most pairs indicate the efficiency of our MC/MD method and suggest a likely large entropy of mixing.

\begin{table}[h]
\begin{tabular}{r|rrrr|rrrr|rrrr}
\multicolumn{1}{c|}{} & \multicolumn{4}{c|}{Swap} &
\multicolumn{4}{c}{NN Bonds} & \multicolumn{4}{c}{NNN Bonds} \\
\hline
$\alpha\backslash\beta$
   &   Zr &   Ti &   Nb &    V &   Zr &   Ti &   Nb &    V &   Zr &   Ti &   Nb &    V \\
\hline
Zr &    - & 0.19 & 0.24 & 0.06 & 0.055& 0.067& 0.060& 0.068& 0.073& 0.060& 0.064& 0.053\\
Ti &      &    - & 0.57 & 0.19 &      & 0.058& 0.065& 0.061&      & 0.061& 0.064& 0.066\\
Nb &      &      &    - & 0.15 &      &      & 0.061& 0.064&      &      & 0.055& 0.067\\
V  &      &      &      &    - &      &      &      & 0.057&      &      &      & 0.064\\
\end{tabular}
\caption{\label{tab:swap-NTVZ} Monte Carlo swap rates and bond counts for Nb-Ti-V-Zr quaternary at $T=1200$K.  Bond counts $N_{\alpha,\beta}$ count number of $\beta$-type neighbors of atom type $\alpha$, where $\alpha$ labels rows and $\beta$ labels columns.  Elements are arranged in order of decreasing BCC lattice constant.}
\end{table}

\begin{table}[h]
\begin{tabular}{r|rrrr|rrrr|rrrr}
\multicolumn{1}{c|}{} & \multicolumn{4}{c|}{Swap} &
\multicolumn{4}{c}{NN Bonds} & \multicolumn{4}{c}{NNN Bonds} \\
\hline
$\alpha\backslash\beta$
   &   Nb &   Mo &    V &   Cr &   Nb &   Mo &    V &   Cr &   Nb &   Mo &    V &   Cr \\
\hline
Nb &    - & 0.44 & 0.23 & 0.05 &  0.049& 0.063& 0.062& 0.076& 0.080& 0.067& 0.056& 0.045\\
Mo &      &    - & 0.50 & 0.21 &       & 0.058& 0.067& 0.062&      & 0.061& 0.064& 0.059\\
 V &      &      &    - & 0.45 &       &      & 0.055& 0.065&      &      & 0.066& 0.066\\
Cr &      &      &      &    - &       &      &      & 0.048&      &      &      & 0.080\\
\end{tabular}
\caption{\label{tab:swap-CMNV} Monte Carlo swap rates and bond statistics for Cr-Mo-Nb-V quaternary at $T=1200$K. Other details as in Table~\ref{tab:swap-NTVZ}.}
\end{table}

\begin{figure}[h]
\includegraphics[clip,width=6.5cm,angle=-90]{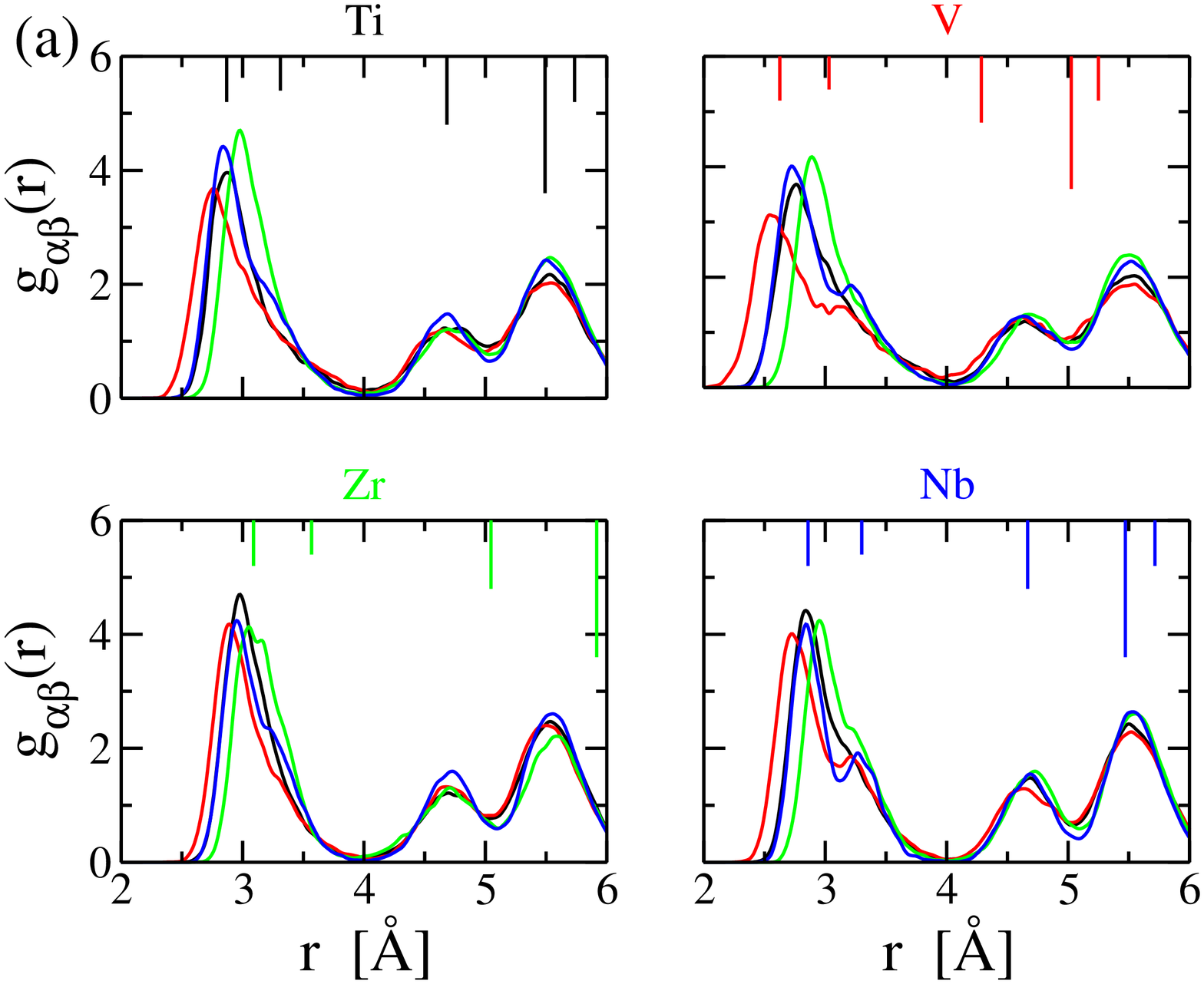}
\includegraphics[clip,width=6.5cm,angle=-90]{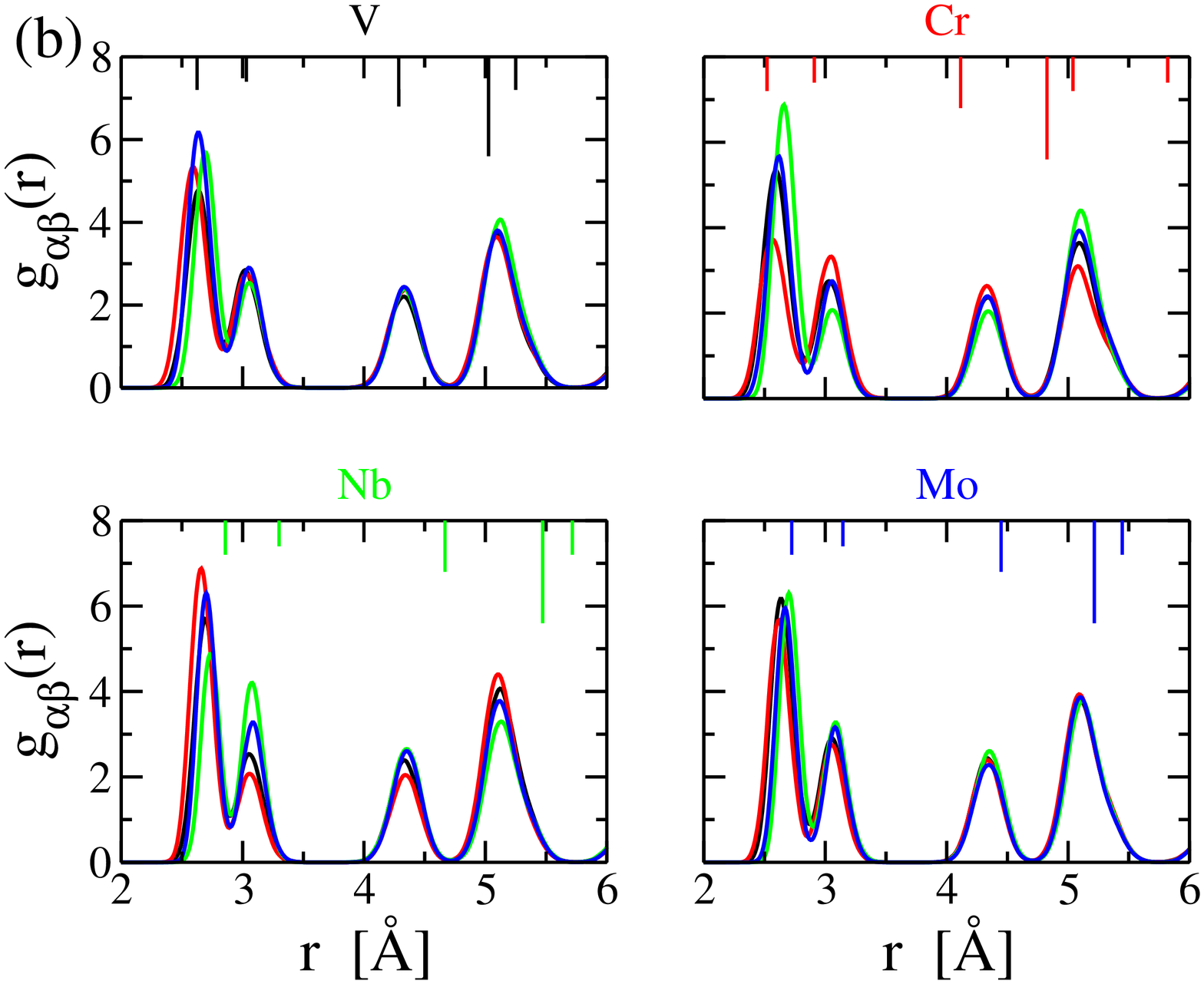}
\caption{\label{fig:pdf} Pair distribution functions of (a) NbTiVZr and (b) CrMoNbV at T=300K quenched from 1200K.  Each panel shows the four partial pair correlation functions for the element named above.  The partials are color coded, {\em e.g.} in (a) for NbTiVZr, black indicates Ti, red indicates V, green indicates Zr and blue indicates Nb.  Thus under Ti the black curve is Ti-Ti and the red curve is Ti-V, {\em etc.}.  Bars at top indicate the corresponding correlations in the pure BCC element, {\em e.g.} elemental Ti has 8 neighbors at $r=2.9$~\AA, 6 at 3.3, 12 at 4.7, 24 at 5.5 and 8 at 5.7.}
\end{figure}

Pair correlation functions are more revealing at lower temperature where atomic motion is reduced.  To mimic an experiment where the alloy is formed at high temperature then cooled down, we quench from T=1200K down to 300K by performing molecular dynamics alone without Monte Carlo swapping. The chemical order remains characteristic of T=1200K even while the phonons equilibrate at 300K, as is the case in typical metallurgical experiments.  Inspecting Fig.~\ref{fig:pdf} note that CrMoNbV exhibits more ideal BCC structure at short range, with the nearest neighbor (NN) and next-nearest neighbor (NNN) peaks clearly resolved, than is the case for NbTiVZr, where these peaks are poorly resolved for Nb and V, and not resolved at all for Zr and Ti.  Nonetheless, the long-range BCC structure is well preserved in both compounds as can be seen in the further neighbor peaks.  A similar local instability was observed in ternary HfNbZr~\cite{Guo13} which likewise mixes elements from the fourth and fifth periodic table columns ({\em i.e.} HEX and BCC ground states).

Because BCC is a loosely-packed structure, instantaneous atomic positions can be mapped back to the ideal BCC lattice sites that they are closest to.  This allows us to uniquely identify NN and NNN bonds and report their statistics, as is done in Tables~\ref{tab:swap-NTVZ} and~\ref{tab:swap-CMNV}.  Notice that the largest number of NN pairs occurs between the elements on the positive diagonals of the squares (Zr-V and Nb-Cr).  This is reflected in Fig.~\ref{fig:pdf} in the respective strengths of the peaks.

\subsection{Diffuse scattering}

\begin{figure}[h]
\hspace{0.2cm}
\includegraphics[trim = 27mm 95mm 25mm 15mm,clip,width=6.5in]{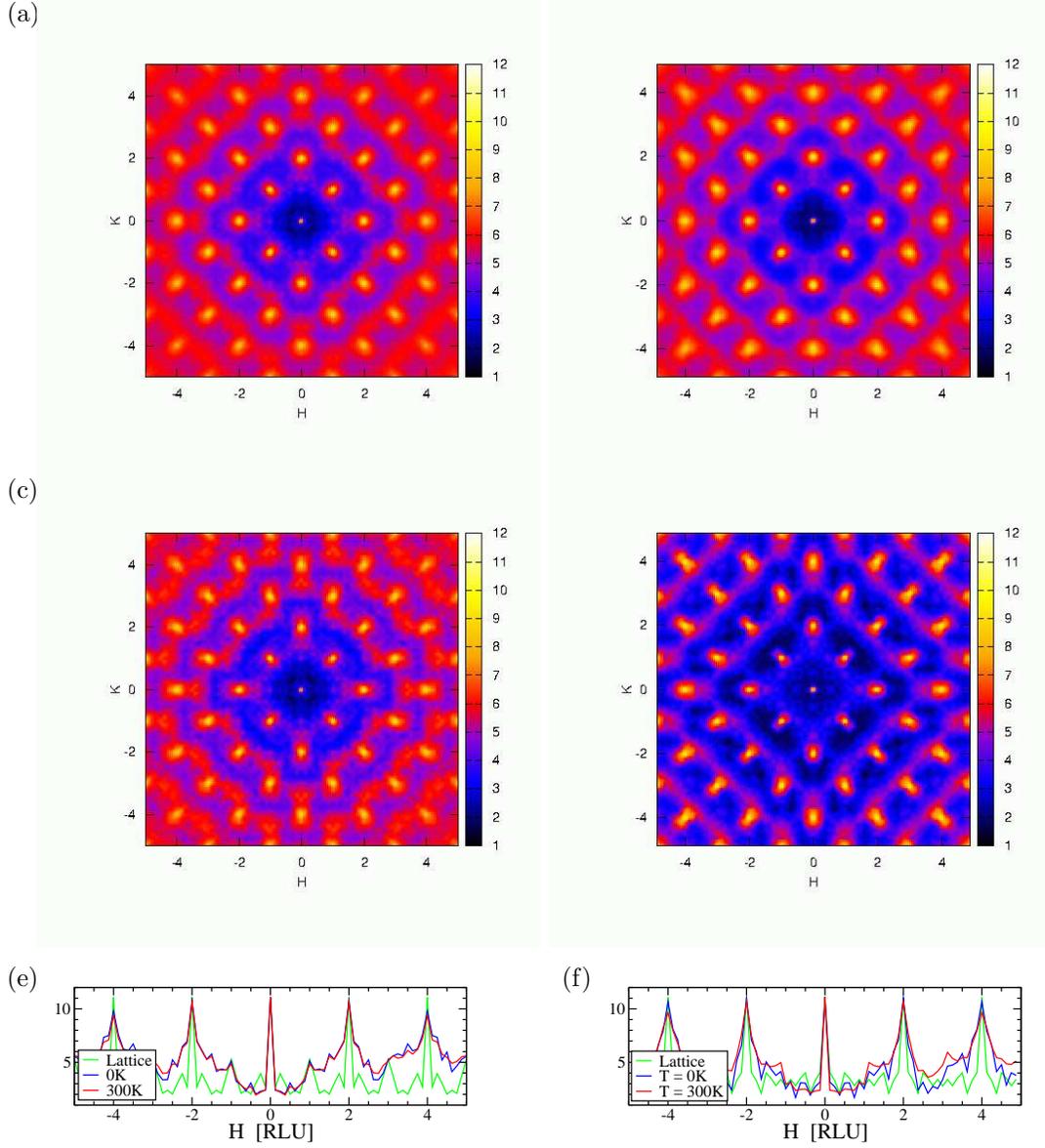}
\caption{\label{fig:diff} Diffuse scattering patterns (natural logarithm of intensity $I(\bQ)$) of NbTiVZr (a, c and e) and CrMoNbV (b, d and f) at temperatures $T=300$K (a and b) and $T=0$K (c and d). Patterns were obtained from samples of 256 atoms in $8\times 8\times 2$ supercells of the BCC unit cell. Curves marked ``Lattice'' in $(H,0,0)$ scans (e and f) arise from placing the atoms at their ideal lattice positions.}
\end{figure}

An instantaneous configuration is described by positions $\br_m$ that are displaced from the ideal lattice sites $\bR_m$. We specify chemical occupation at position $\br_m$ by $c_m^\alpha$ defined as 1 if occupied by species $\alpha$ and 0 otherwise. The scattering amplitude at reciprocal space position $\bQ$ is~\cite{Borie57,Krivoglaz96,Schweika98,Welberry04}
\begin{equation}
A(\bQ)=\sum_m \sum_\alpha f_\alpha c_m^\alpha e^{-i\bQ\cdot\br_m}.
\end{equation}
and its square is the scattering intensity $I(\bQ)=|A(\bQ)|^2$.  Here $f_\alpha$ is the scattering form factor $f_\alpha$. For the present analysis we utilize x-ray form factors which we approximate as the atomic number, $f_m\approx Z_m$, independent of $\bQ$.  Note that the set of atomic numbers present in NbTiVZr (22, 23, 40 and 41) exhibits contrast similar to the set present in CrMoNbV (23, 24, 41, and 42).  The natural logarithm of the intensity is plotted in Fig.~\ref{fig:diff}, which is measured in reciprocal lattice units (RLU), $(2\pi/a)$, where $a$ is the conventional cubic lattice constant.  Bragg peaks of the BCC lattices occur at $\bQ=\bG=(2\pi/a)(H,K,L)$ where $H, K$ and $L$ are integers such that $H+K+L$ is even.  For $\bQ$ close to a Bragg peak $\bG$ we shall denote the deviation by $\bk=\bQ-\bG$.

The average scattering intensity $I(\bQ)=\avg{|A(\bQ)|^2}$ separates into the Bragg component of the average lattice $I_B(\bQ)=|\avg{A(\bQ)}|^2$, and the diffuse part due to fluctuations $I_D(\bQ)=\avg{|A(\bQ)-\avg{A(\bQ)}|^2}=I-I_B$.  Owing to configurational fluctuations we take averages over both atomic displacements $\bu_m=\br_m-\bR_m$ and chemical species fluctuations $\Delta c_m^\alpha=c_m^\alpha-c^\alpha$, yielding~\cite{Schweika98}
\begin{equation}
\label{eq:ID}I_D(\bQ)=
\sum_{mn} e^{-i\bQ\cdot(\bR_{mn})} \sum_{\alpha\beta} \avg{f_\alpha f_\beta 
\Delta c_m^\alpha \Delta c_n^\beta e^{-i\bQ\cdot(\bu_{mn})}}
\end{equation}
where we introduce the relative coordinates $\bR_{mn}=\bR_n-\bR_m$ and $\bu_{mn}=\bu_n-\bu_m$.
For sufficiently small $\bQ\cdot\bu$, we may expand the exponential
\begin{equation}
e^{-i\bQ\cdot(\bu_{mn})} \approx
1-i\bQ\cdot(\bu_{mn})-\frac{1}{2}(\bQ\cdot(\bu_{mn}))^2+\cdots
\end{equation}
Separating the diffuse scattering into components, we set $I_D=I_0+I_1+I_2+\cdots$, where each term arises from successively higher powers of $\bQ\cdot\bu$. Here $I_0$ reflects short-range chemical order, $I_1$ is known as the size effect, and $I_2$ includes both the thermal diffuse scattering due to lattice vibrations (phonons) and also the Huang scattering due to correlated lattice strains caused by differing atomic sizes.

To model chemical disorder, assume that all atoms sit on BCC lattice sites $\{\bR_m\}$ with chemical species randomly distributed.  For $\bQ$ not on a Bragg peak, i.e. $\bk=\bQ-\bG\neq 0$, the sum over $m$ and $n$ in Eq.~(\ref{eq:ID}) simplifies to $I_0/N\sim \avg{f^2}-\avg{f}^2$ independent of $\bk$, resulting in a uniform diffuse background.  However, our neighbor statistics given in Tables~\ref{tab:swap-NTVZ} and~\ref{tab:swap-CMNV} show the distributions of species is {\em not} fully random, but rather exhibits short-range correlations.  Unlike species (e.g. Zr with V in NbTiVZr, or Nb with Cr in CrMoNbV) prefer to occupy nearest neighbor bonds.  On a BCC lattice such sort-range chemical order can lead to partial B2-like ordering in which broad diffuse maxima arise at previously forbidden positions where $H+K+L$ is odd, as indeed we see in the case of NbTiVZr (Fig.~\ref{fig:diff}(a)).  Despite the short-range order present in CrMoNbV, we see no B2-like peaks in its diffraction pattern, which could indicate more complex order such as the B2$_3$ predicted in NbMo~\cite{Blum05}. 

In fact, the atoms in our models do not sit at ideal lattice sites.  The random distribution of unlike atomic sizes cause substantial relaxations off the lattice sites even at T=0K, and in addition they are subject to ordinary thermal fluctuations. Atomic size effects enter both at first and second order.  At the first order the ``atomic size effect'' yields~\cite{Welberry04}
\begin{equation}
I_1(\bQ) = - \sum_{\alpha\beta} f_\alpha f_\beta \sum_{mn}\avg{\Delta c_m^\alpha \Delta c_n^\beta \bQ\cdot \bu_{mn}} \sin{\bQ\cdot\bR_{mn}}.
\end{equation}
Nonvanishing $I_1$ depends on a correlation between the sign of $\bu_{mn}$ and the strengths of the form factors at sites $m$ and $n$.  The long-range displacement $\bu_{mn}\sim \be_{\bR}/R^2$ has Fourier transform $\bu(\bk)\sim \be_{\bk}/k$.  Thus close to a Bragg peak at $\bQ=\bG+\bk$, the size effect scattering diverges~\cite{Schweika98} as $I_1(\bQ)\sim-\bG\cdot\be_\bk/k$ and will be asymmetric on the two sides of $\bG$. For example, if atoms with large $f_\alpha$tend to move apart and small $f_\alpha$ tend to move closer, then $I_1$ is positive for negative $\bk$ and negative for positive $\bk$~\cite{Welberry04}.  This asymmetry can give the impression of a shift in the peak position~\cite{Warren51}.

At second order ordinary thermal diffuse scattering due to phonons, and also Huang scattering due to atomic sizes, both contribute diffuse scattering that diverges as $1/k^2$ in the vicinity of the Bragg peaks, each with its own characteristic angular distribution.  Ordinary thermal fluctuations are characterized by phonons whose dispersion relations are of the form $\omega_j=c_j k$, where $j=1..3$ enumerate the independent modes of vibration whose polarization vectors are $\be_j(\bk)$.  The resulting diffuse intensity is proportional to~\cite{Egami13}
\begin{equation}
\label{eq:TDS}
\langle|\bG\cdot\bu_\bk|^2\rangle \sim
\sum_j \frac{|\bG\cdot\be_j(\bk)|^2k_BT}{c_j^2k^2}
\end{equation}
As transverse sound speeds are typically less than longitudinal, the diffuse scattering will usually be stronger in directions $\bk$ perpendicular to $\bG$ than parallel to it. Such elongations can be seen in the diffuse pattern for CrMoNbV in surrounding the (200) and (400) peaks. Also visible are streaks running in [110] directions similar to those observed in other BCC metals~\cite{Ramsteiner08}.  These could indicate a low frequency phonon mode with a polarization component in the [110] direction.  Notice that the streaks are stronger on the low $\bQ$ side of the Bragg peaks ({\em i.e.} negative $\bG\cdot\be_\bk$) than on the high $\bQ$ side, presumably as a result of the asymmetry of the size effect scattering $I_1$.  In fact, there are local maxima of scattering in CrMoNbV that shift from, {\em e.g.} $H=3.1$ at $T=300$K to $H=3.5$ at $H=0$K, as can be seen in Fig.~\ref{fig:diff}f indicating the diminishing contribution of the thermal component of $I_2$ relative to $I_1$.

Huang scattering arises from long-range strains arising from atomic size mismatch and hence is temperature independent.  Originally derived for dilute impurities, a similar effect is present in concentrated alloys such as high entropy alloys.  Because these strains are primarily longitudinal, only the component of $\bk$ parallel to $\bG$ enters, and the diffuse intensity is proportional to
\begin{equation}
\label{eq:Huang}
\left(\frac{\bG\cdot\be_\bk}{k}\right),
\end{equation}
which is equivalent to Eq.~\ref{eq:TDS} setting $\be_j$ parallel to $\bk$.  The Huang scattering consists of figure-eights that are characteristically elongated in the longitudinal (radial) direction in reciprocal space.  The low T patterns in Fig.~\ref{fig:diff} indeed show such elongation, and again the asymmetry in the sign of $\bk$ arises from the contribution of $I_1$. The figure-eight pattern is not clearly resolved in the case of NbTiVZr, due to transverse strain that might indicate the presence of multisite correlations.

\subsection{Configurational entropy}

\def\CVMone{{\{1\}}}

\def\CVMdot{{\{\tikz{\filldraw (0,0) circle (0.05cm);}\}}}

\def\CVMNN{{
\{\tikz[baseline]{
\draw (0,-.05)--(.2,.15);
\filldraw (0,-.05) circle (0.05cm);
\filldraw (.2,.15) circle (0.05cm);}\}}}

\def\CVMNNN{{
\{\tikz{\draw (0,0)--(.4,0);
\filldraw (0,0) circle (0.05cm);
\filldraw (.4,0) circle (0.05cm);}\}}}

\def\CVMTriang{{
\{\tikz[baseline]{
\draw (0,-.05)--(.2,.15);
\draw (0,-.05)--(.4,-.05);
\draw (.4,-.05)--(.2,.15);
\filldraw (0,-.05) circle (0.05cm);
\filldraw (.2,.15) circle (0.05cm);
\filldraw (.4,-.05) circle (0.05cm);}\}}}

\def\CVMTetra{{
\{\tikz[baseline]{
\draw (0,.05)--(.2,.25);
\draw (0,.05)--(.2,-.15);
\draw (.4,.05)--(.2,.25);
\draw (.4,.05)--(.2,-.15);
\draw (.2,-.15)--(.2,.25);
\draw (0,0.05)--(0.15,0.05);
\draw (0.25,0.05)--(0.4,0.05);
\filldraw (0,0.05) circle (0.05cm);
\filldraw (.2,.25) circle (0.05cm);
\filldraw (.2,-.15) circle (0.05cm);
\filldraw (.4,.05) circle (0.05cm);}\}}}

We evaluate the configurational entropy using the approach of the cluster variation method (CVM~\cite{deFontaine79,deFontaine94}) by starting with the mean field (Bragg-Williams) entropy, then modifying it through the inclusion of factors that reduce the entropy by correcting for local correlations.  First we define numerical values for symbols associated with the empty lattice, isolated points (P), near-neighbor bonds (NN), next-nearest-neighbor bonds (NNN), triangles (TRI) and tetrahedra (TET) as
\begin{equation}
\label{eq:CVMdef}
\CVMone \equiv M!,~~
\CVMdot\equiv\prod_{\alpha} (x_{\alpha} M)!,~~
\CVMNN\equiv\prod_{\alpha\beta} (y_{\alpha\beta} M)!,~~ 
\end{equation}
\begin{equation}
\CVMNNN\equiv\prod_{\alpha\beta} (w_{\alpha\beta} M)!,~~
\CVMTriang\equiv\prod_{\alpha\beta\gamma} (v_{\alpha\beta\gamma} M)!,~~
\CVMTetra\equiv\prod_{\alpha\beta\gamma\delta} (z_{\alpha\beta\gamma\delta} M)!,
\end{equation}
where $M$ is the total number of lattice sites, $x_\alpha$ are the fractions of species $\alpha$, $y_{\alpha\beta}$ is the frequency of NN bonds between species $\alpha$ and $\beta$, {\em etc.}  Next, we introduce combinatorial factors
\begin{equation}
\label{CVMW}
\Omega_P = \frac{\CVMone}{\CVMdot},~~
\Omega_{NN} = \frac{\CVMdot^8}{\CVMNN^4\CVMone^4},~~
\Omega_{NNN} = \frac{\CVMdot^{6}}{\CVMNNN^3\CVMone^3},~~
\Omega_{TET} = \frac{\CVMTriang^{12}\CVMone^6}{\CVMTetra^6\CVMdot^{12}}.
\end{equation}
Note that $\frac{1}{M}\log{\Omega_P}\approx-\sum_\alpha x_\alpha\log{x_\alpha}$ is the mean field entropy~\cite{Bragg34}.  Multiplying $\Omega_P$ by $\Omega_{NN}$ reduces the entropy by an amount related to the deviation of NN bond frequencies $y_{\alpha\beta}$ from the uncorrelated frequency $x_\alpha x_\beta$.  This level of approximation also goes by the name ``quasichemical approximation''~\cite{Guggenheim44}.  Multiplying instead by $\Omega_{NNN}$ would do the same for the NNN bonds, so the product $\Omega_{NN}\Omega_{NNN}$ given in Table~\ref{tab:ent} incorporates both effects, that is, it corrects the quasichemical approximation through inclusion of NNN correlations.  Finally, including the remaining factor $\Omega_{TET}$ results in the conventional CVM entropy expression for BCC lattices~\cite{Ackermann89}.

\begin{table}
\begin{tabular}{l|ccccc}
$\exp{(S/k_B)}$ & MF & QC    &    NNN & CVM \\
\hline
NbTiVZr         & 4  & 3.971 &  3.931 & (3.789) \\
CrMoNbV         & 4  & 3.886 &  3.759 & (3.657) \\
\end{tabular}
\caption{\label{tab:ent} Entropies of NbTiVZr and CrMoNbV at $T=1200$K in the mean field approximation (MF~\cite{Bragg34}), the quasichemical approximation (QC~\cite{Guggenheim44}), an improved QC incorporating both NN and NNN correlations, and the cluster variation (CVM) method~\cite{Ackermann89}.  CVM entropies are poorly converged owing to limited sampling statistics of the four-point correlation. Values quoted are phase space volumes $\exp{(S/k_B)}$ where $S$ is the entropy per site.}
\end{table}


By accumulating cluster occupation statistics during our MC/MD simulation, we are able to evaluate the numerical factors.  Table~\ref{tab:ent} presents the resulting entropies (in the form of the phase space volume per site, which is the exponential of the entropy per site) for our two compounds. Note that these are configurational entropies only ({\em i.e.} no vibrational or electronic contributions), and that they reflect the chemical disorder at T=1200K.  The entropies decrease monotonically as additional correlations are included.  However, the CVM entropy must not be taken as an upper bound, because our sampling statistics are insufficient for the four point correlation function $z_{\alpha\beta\gamma\delta}$, and additional run time would increase this value. The proper values lie between NNN and CVM, and we recommend taking the NNN value as our best estimate.  Interestingly, the entropy of NbTiVZr lies closer to the ideal (mean field) value than CrMoNbV does.  This might reflect the extreme short-range disorder of NbTiVZr, which seemingly precludes establishing strong local chemical order.

\section{Conclusion}

In summary, we have modeled the structures of two body-centered cubic refractory high entropy alloys, NbTiVZr and CrMoNbV. We discover a local instability of NbTiVZr that reduces the short-range BCC structure while preserving the long-range lattice. This instability is likely due to the preference of Ti and Zr for HEX structures at low temperature. Chemical disorder, atomic size differences and thermal fluctuations lead to diffuse x-ray scattering patterns, with chemical and size effects dominating at room temperature in the case of NbTiVZr, while thermal fluctuations dominate in the case of CrMoNbV. A hint of B2-like chemical order is evident in NbTiVZr but is surprisingly absent in CrMoNbV. A side benefit of our simulations is the ability to evaluate configurational entropy from the pair correlation functions. We find the entropies at $T=1200$K are $k_B\ln{(3.93)}$ for NbTiVZr and $k_B\ln{(3.76)}$ for CrMoNbV.

\section*{Acknowledgements}
I thank Michael Gao, Marcel Sluiter, Soumyadipta Maiti and Walter Steurer for useful discussions.

\bibliography{refs}

\clearpage

\end{document}